\documentclass{emulateapj}
\usepackage{epsfig}
\usepackage{color}

\newcommand   {\about} {\mbox{$\sim$}}

\newcommand   {\hh}    {\mbox{H$_2$}}

\newcommand   {\nhhh}  {\mbox{NH$_3$}}
\newcommand   {\fivnhhh}  {\mbox{$^{15}$NH$_3$}}
\newcommand   {\founhhh}  {\mbox{$^{14}$NH$_3$}}

\newcommand   {\nhhd}  {\mbox{NH$_2$D}}

\renewcommand {\deg}   {\mbox{$^\circ$}}
\newcommand   {\arcm}  {\mbox{$^\prime$}}
\newcommand   {\arcs}  {\mbox{$^{\prime\prime}$}}

\newcommand   {\pccm}  {\mbox{cm$^{-3}$}}

\newcommand   {\kms}   {\mbox{km\,s$^{-1}$}}
\renewcommand {\ga}    {\mbox{\rlap{\hbox{\lower4pt\hbox{$\sim$}}}\hbox{$>$}}}
\renewcommand {\la}    {\mbox{\rlap{\hbox{\lower4pt\hbox{$\sim$}}}\hbox{$<$}}}

\newcommand   {\hr}    {\mbox{$^{\rm h}$}}
\newcommand   {\mt}    {\mbox{$^{\rm m}$}}
\newcommand {\secper}  {\mbox{\rlap{\hbox{\hbox{.}}}\hbox{$^{\rm s}$}}}

\received{2009 November 16}
\accepted{{2009 December 30}\\*[0.06in] }

\slugcomment{To appear in {The Astrophysical Journal Letters}}

\shorttitle{Nitrogen Fractionation}
\shortauthors{Lis et~al.}

\begin{document}

\title{Nitrogen Isotopic Fractionation in Interstellar Ammonia}

\author{D.~C.~Lis\altaffilmark{1},
A.~Wootten\altaffilmark{2},
M.~Gerin\altaffilmark{3}, and
E.~Roueff\altaffilmark{4}
}

\altaffiltext{1}{California Institute of Technology, Cahill Center for
  Astronomy and Astrophysics 301-17, Pasadena, CA~91125; dcl@caltech.edu} 

\altaffiltext{2}{National Radio Astronomy Observatory, 520 Edgemont
  Road, Charlottesville, VA~22903; awootten@nrao.edu}

\altaffiltext{3}{LERMA, CNRS UMR8112, Observatoire de Paris and
Ecole Normale Superieure, 24 Rue Lhomond, 75231 Paris cedex 05,
France; maryvonne.gerin@lra.ens.fr}

\altaffiltext{4}{LUTh, CNRS UMR8102, Observatoire de Paris,
Section de Meudon, Place J. Janssen, 92195 Meudon, France;
evelyne.roueff@obspm.fr}

\begin{abstract}
%\vspace {5pt}
  Using the Green Bank Telescope (GBT), we have obtained accurate
  measurements of the $^{14}$N/$^{15}$N isotopic ratio in ammonia in
  two nearby cold, dense molecular clouds, Barnard~1 and NGC~1333. The
  $^{14}$N/$^{15}$N ratio in Barnard~1, $334 \pm 50$ (3$\sigma$), is
  particularly well constrained and falls in between the local
  interstellar medium/proto-solar value of \about 450 and the
  terrestrial atmospheric value of 272. The NGC~1333 measurement is
  consistent with the Barnard~1 result, but has a larger uncertainty.
  We do not see evidence for the very high $^{15}$N enhancements seen
  in cometary CN.
% and predicted by chemical models to be present in
%  cold, dense, CO depleted regions. 
  Sensitive observations of a
  larger, carefully selected sample of prestellar cores with varying
  temperatures and gas densities can significantly improve our
  understanding of the nitrogen fractionation in the local
  interstellar medium and its relation to the isotopic ratios measured
  in various solar system reservoirs.
\end{abstract}

\keywords{astrochemistry --- ISM: abundances --- ISM: individual
  (Barnard~1, NGC~1333) --- ISM: molecules --- molecular processes ---
  radio lines: ISM}

\section{Introduction}\label{sec:introduction}

Molecular isotopic ratios are an important tool in the study of the
origin of interstellar and solar system materials. After hydrogen,
nitrogen displays the largest isotopic variations in the solar system,
typically explained by mixing of various proto-solar or pre-solar
reservoirs. Earth, Mars interior, Venus and most primitive meteorites
have nitrogen isotopic ratios within 5\% of the terrestrial
atmospheric value, $^{14}$N/$^{15}$N~$= 272$ \citep[see][]{marty09}.
However, the proto-solar nebula was poorer in $^{15}$N, with
$^{14}$N/$^{15}$N~$\simeq 450$, as evidenced by infrared and in-situ
measurements in the Jupiter atmosphere ($530 ^{+380}_{-170}$, ISO,
\citealt{fouchet00}; $435 \pm 57$, Galileo, \citealt{owen01}; $448 \pm
62$, Cassini, \citealt{abbas04}; $450 \pm 106$, Cassini,
\citealt{fouchet04}) and recent solar wind measurements ($442 \pm
131$, 2$\sigma$, Genesis, \citealt{marty09}). The proto-solar
$^{14}$N/$^{15}$N ratio is in agreement with the local interstellar
medium (ISM) value ($450 \pm 22$, \citealt{wilson94}; or $414 \pm
32$ at the birth place of the Sun, \citealt{wielen97}).

Very low nitrogen isotopic ratios, $^{14}$N/$^{15}$N~$= 148 \pm 6$,
have been measured in CN in a large sample of comets
\citep{manfroid09}. Similar isotopic ratios have been recently derived
by \cite{bockelee08} in comet 17P/Holmes in CN and HCN, the presumed
parent of the CN radical in cometary atmospheres ($^{14}$N/$^{15}$N~$=
165 \pm 40$ and $139 \pm 26$, respectively). Although higher ratios in
HCN were initially reported in comet Hale-Bopp, Bockel\'{e}e-Morvan's
re-analysis of the archival data suggests that the $^{14}$N/$^{15}$N
ratio in HCN in this comet is also consistent with the CN value, given
the measurement uncertainties. The exact chemical networks explaining
the HCN isotopic anomaly in comets have yet to be proposed. However,
one clear conclusion is that HCN has never isotopically equilibrated
with the nebular N$_2$ at the later phases of the evolution of the
solar system \citep{bockelee08}. Low nitrogen isotopic ratios are also
found in some meteorites, interplanetary dust particles (IDPs) and
samples of material returned from comet 81P/Wild 2 by the
Stardust spacecraft; amine groups in particular often show low
$^{14}$N/$^{15}$N ratios \citep{engel97, keller04, floss06, mckeegan06}.

In the ISM, nitrogen bearing molecules are important tracers of cold,
dense gas, as they do not freeze-out onto grain mantles at densities
below a few $\times 10^6$~\pccm, contrary to carbon and oxygen
species, such as CO or CS. In cold, dense, CO depleted ISM regions,
high $^{15}$N enhancements have been suggested to be present in
gas-phase molecules, in particular in ammonia \citep[see][and
references therein]{rodgers08a, rodgers08b}. These isotopic
enhancements could be preserved as spatial heterogeneity in ammonia
ice mantles, as monolayers deposited at different times would have
different isotopic compositions and may in turn explain the $^{15}$N
enhancements seen in primitive solar system refractory organics, if
they are synthesized from ammonia.

Measurements of nitrogen isotopic ratios in ammonia in cold molecular
clouds are of great interest, as they provide quantitative
observational constraints for the chemical fractionation models.
However the data available are scarce. In a recent study,
\cite{gerin08} detected a doubly-fractionated ammonia isotopologue,
$^{15}$NH$_2$D, in several ISM regions. They derive $^{14}$N/$^{15}$N
isotopic ratios \about 350--850, similar to the proto-solar
value of 450. However, the $^{15}$NH$_2$D lines are weak and the
corresponding measurement uncertainties are large. To better constrain
the nitrogen isotopic ratios in the solar neighborhood, we have
observed the \founhhh\ and \fivnhhh\ inversion lines using the NRAO
100~m Robert C. Byrd Green Bank Telescope. We present here results for
two nearby clouds, Barnard~1 and NGC~1333, and discuss prospects for
future studies of nitrogen isotopic ratios using the GBT.

\section{Observations}\label{sec:observations}

\begin{figure}[tb]
\centering
\includegraphics[height=0.95\columnwidth,angle=270]{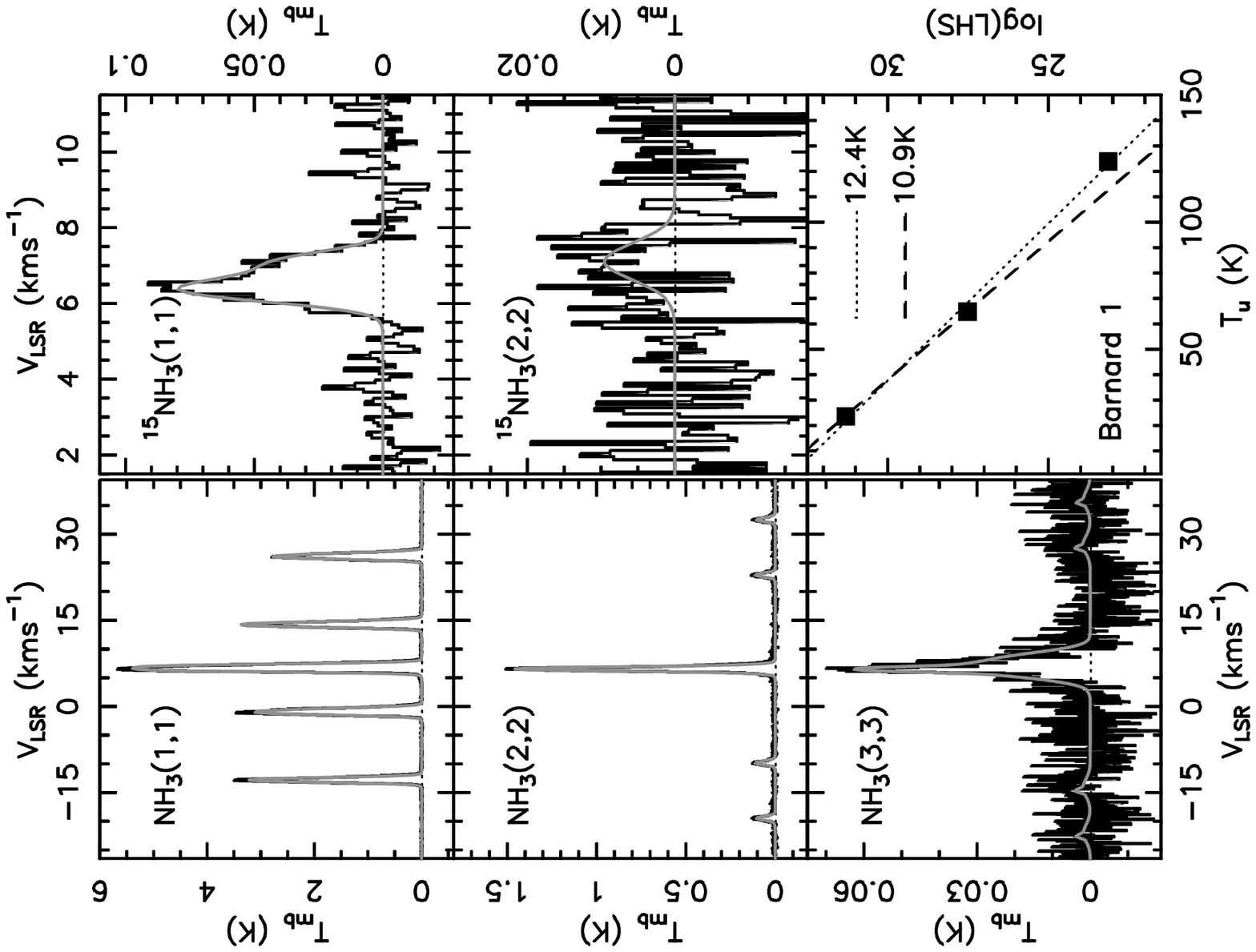}
\caption{Spectra of the ammonia inversion lines in Barnard~1-b
  ($\alpha_{J2000}$ = 03\hr 33\mt 20\secper 8, $\delta_{J2000}$ =
  +31\deg 07\arcm 34\arcs). Gray lines show hyperfine structure
  fits---a single velocity component for the \founhhh~(1,1), (2,2),
  and \fivnhhh~(1,1) transitions and two components for the
  \founhhh~(3,3) transition, as described in the text. The
  \fivnhhh~(2,2) spectrum, shown with a Gaussian line fit, is formally
  a 3.5$\sigma$ upper limits. The bottom-right panels shows the LTE
  rotation diagram, as described in the text. LHS stands for the
  left-hand-side in eq. (2) of Blake et al. (1987). }
\label{fig:spec1}       % Give a unique label
\end{figure}

\begin{figure}[tb]
\centering
\includegraphics[height=0.95\columnwidth,angle=270]{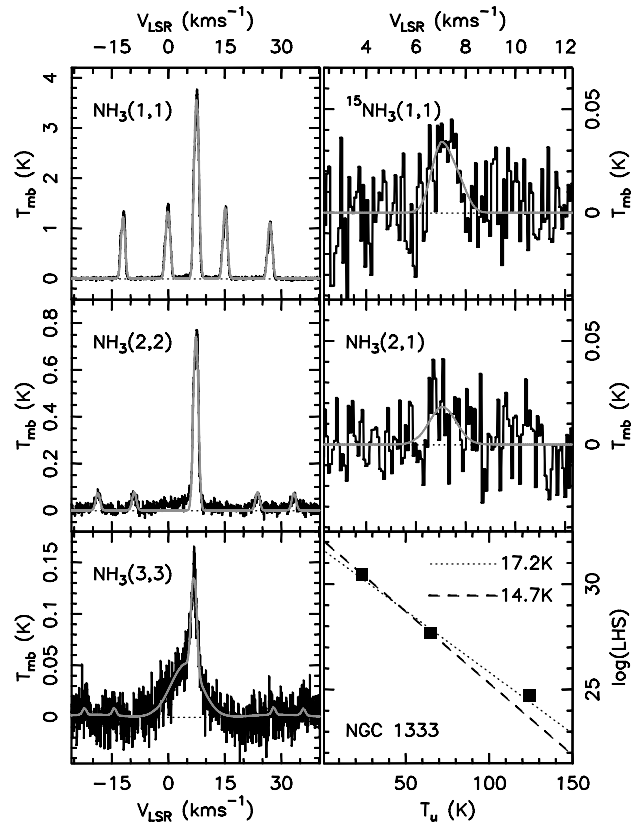}
\caption{Spectra of the ammonia inversion lines in NGC~1333
  ($\alpha_{J2000}$ = 03\hr 29\mt 11\secper 6, $\delta_{J2000}$ =
  +31\deg 13\arcm 26\arcs). The \founhhh~(2,1) spectrum is formally a
  3.7$\sigma$ upper limit. The bottom-right panel shows the LTE
  rotation diagram, as described in the text. Lines as in Fig.~1.}
\label{fig:spec2}       % Give a unique label
\end{figure}

Observations of ammonia inversion lines presented here were carried
out in 2009 February using the GBT. Conditions were calm under a
lightly overcast sky resulting in a stable system temperature of
\about 48--50~K. The K-band receiver feed/amplifier set covering
22--26.5~GHz was employed. The GBT spectrometer was configured in its
eight 50~MHz intermediate frequency (IF) bandwidth, nine-level mode,
which allows simultaneous coverage of four 50~MHz frequency bands in
two polarizations, through the use of offset oscillators in the IF.
This mode gives 3.051~kHz channel separation. The IF No. 0 was
centered between the (1,1) and (2,2) lines of \founhhh, the (3,3) line
was centered in IF No.~1 and the (2,1) line in IF No.~3; IF No.~2 was
tuned midway between the (1,1) and (2,2) lines of $^{15}$NH$_3$.

The antenna temperature calibration was achieved by injecting a
calibrated, broadband signal from a noise diode. The data were placed
on the $T_A^*$ scale \citep{ulich76} by multiplying by a factor of
e$^{\tau A}/ \eta_l$, where $\tau$ is the zenith opacity (\about 0.065
at 23.5~GHz and 0.075 at 22.5~GHz, from an atmospheric model), $A$ is
the air mass of the observations, and $\eta_l$ is the rear spillover
efficiency (\about 0.99 for the unblocked GBT aperture). The source
0336+3218 was used for pointing and absolute calibration at the
beginning of the observation set. The absolute calibration accuracy is
conservatively estimated at 20\%. Relative accuracy among lines in
spectral bandpasses measured at the same time is much better,
typically dominated by the random noise of the spectra. The
conventional beam efficiency of the GBT at 20~GHz is $\eta_{mb} \simeq
0.79$ and the FWHM beam size is \about 33\arcs. Data were taken in the
frequency switching mode, with a frequency displacement of 6 MHz.

Automatically updated dynamic pointing and focusing corrections were
employed based on real-time temperature measurements and a thermal
model of the GBT; zero points were adjusted typically every 2 hours or
less, using 0336+3218 as a calibrator. The two polarization outputs
from the spectrometer were averaged in the final data reduction
process to improve the signal-to-noise ratio. The total integration
times for Barnard~1 and NGC~1333 are 174 and 48 min, respectively.

\section{Results}\label{sec:results}

\begin{table*}[tb]       
\caption{HFS Fit Results and Ammonia Column Densities.}\label{tab:hfsfits}
\begin{center}
%\scriptsize
\begin{tabular}{lccccc}
\noalign{\vskip -3.0mm}
\hline
\hline
\noalign{\smallskip}
Transition & $T_R^* dv^a$           & $V_{LSR}$         & $\Delta V$    & 
$\tau$ & $N_{mol}$     \\ 
                  &  (K km s$^{-1}$)  & (km s$^{-1}$) & (km s$^{-1}$)  &
            & (cm$^{-2}$)   \\
\hline
\hline
\noalign{\smallskip} 
\multicolumn{6}{c}{Barnard 1-b}\\
\hline
\noalign{\smallskip}
% Line     T ant * Tau           V lsr          delta V            Tau main
%  1   22.2   (2.825E-4)   6.69   (1.659E-6)  0.797   (1.320E-4)   3.98   (4.188E-5)
%   Area=2.000*22.2*0.797*SQRT(PI()/4/LN(2))=37.67; 
%  error = 37.67*sqrt((2.825E-4|22.2)^2+(1.320e-4|0.797)^2) = 6.26e-3
\founhhh~(1,1) & $37.7 \pm 0.006$ & 6.7  & 0.80 & 4.0 & $2.8 \times 10^{15} $\\
% Line     T ant * Tau           V lsr          delta V            Tau main
%   1   1.79   (3.842E-3)   6.53   (3.908E-4)  0.795   (1.265E-3)  0.292   (5.031E-3)
% Area=1.255812*1.79*0.795*SQRT(PI()/4/LN(2))=1.902
% error = 1.902*sqrt((3.842E-3|1.79)^2+(1.265E-3|0.795)^2) = 5.08e-3
\founhhh~(2,2) & $1.90 \pm 0.005$  & 6.5 & 0.80 & 0.3 & $2.8 \times 10^{15} $ \\

% Line     T ant * Tau           V lsr          delta V            Tau main
%   1  6.077E-2(2.463E-2)   7.11   (0.102   )   3.46   (0.472   )   1.65   ( 1.18   )
% Area=1.14255*6.077e-2*3.46*SQRT(PI()/4/LN(2))=0.256
% error = 0.256*sqrt((2.463E-2|6.077E-2)^2+(0.472|3.46)^2) = 0.109
%   2  7.812E-2(2.967E-2)   6.46   (5.168E-2)  0.796   ( 0.00   )   1.92   ( 1.15   )
% Area=1.14255*7.812e-2*0.796*SQRT(PI()/4/LN(2))=0.0756
% error = 0.0756*sqrt((2.967E-2|7.812E-2)^2+(0|0.796 )^2) = 0.0287
\founhhh~(3,3) & $0.076 \pm 0.029 $ & 6.5 & 0.80$^b$ & ---& $8.0 \times 10^{15} $ \\
                         & $0.26 \pm 0.11$ & 7.1 & 3.5          &
                         --- & --- \\
\founhhh~(2,1) & $0.012 \pm 0.004$ & --- & --- & --- & --- \\

% Line     T ant * Tau           V lsr          delta V            Tau main
%   1  0.104   (6.362E-3)   6.62   (2.542E-2)  0.885   (6.146E-2)  0.100   (0.482   )
%   Area=0.104*0.885*SQcRT(PI()/4/LN(2))=9.80e-2; error 9.07e-3
%\fivnhhh~(1,1) &  $0.098 \pm 0.004$ & 6.6 & 0.89 &
%--- & $7.8 \times 10^{12} $ \\
%   1  0.158   (5.131E-2)   6.59   (2.134E-2)  0.629   (8.555E-2)  0.794   ( 1.06   )
\fivnhhh~(1,1)$^c$ &  $0.106 \pm 0.004$ & 6.6 & 0.63 &
--- & $8.4\times 10^{12} $ \\
% Line      Area                   Position            Width             Tpeak
%  1    1.16760E-02 (  0.003)    7.108 (  0.226)    1.150 (  0.000)  9.53818E-03
\fivnhhh~(2,2) &  $0.013 \pm 0.004$ & --- & --- & --- & --- \\
\hline
\noalign{\smallskip} 
\multicolumn{6}{c}{NGC~1333}\\
\hline
\noalign{\smallskip}
% Line     T ant * Tau           V lsr          delta V            Tau main
%   1   6.38   (4.598E-3)   7.48   (5.468E-4)   1.21   (6.811E-4)   1.21   (3.047E-3)
%   Area=2.000*6.38*1.21*SQRT(PI()/4/LN(2))=16.43
% error = 16.43*sqrt((4.598E-3|6.38)^2+(6.811E-4|1.21)^2) = 0.0150
\founhhh~(1,1) & $16.4 \pm 0.015$ & 7.5  & 1.2 & 1.2 & $8.9 \times 10^{14} $\\
% Line     T ant * Tau           V lsr          delta V            Tau main
%  1   1.25   (3.745E-2)   7.31   (5.842E-3)   1.33   (1.616E-2)   1.10   (8.238E-2)
% Area=1.255812*1.25*1.33*SQRT(PI()/4/LN(2))=2.22
% error = 2.22*sqrt((3.745E-2|1.25)^2+(1.616E-2|1.33)^2) = 0.072
\founhhh~(2,2) & $2.22 \pm 0.07 $ & 7.3 & 1.3 & 1.1 & $9.9 \times 10^{14} $\\
% Line     T ant * Tau           V lsr          delta V            Tau main
%   1  5.301E-2(2.488E-3)   4.71   (0.187   )   9.52   (0.410   )  0.100   (0.141   )
% Area=1.14255*5.301e-2*9.52*SQRT(PI/4/LOG(2))=0.614
% error = 0.614*sqrt((2.488E-3 |5.301E-2 )^2+( 0.410 |9.52  )^2) = 0.039
%   2  0.236   (4.070E-2)   6.80   (4.232E-2)   1.28   ( 0.00   )   2.37   (0.631   )
% Area=1.14255*0.236*1.28*SQRT(PI/4/LOG(2))=0.367
% error = 0.367*sqrt((4.070E-2|0.236)^2+(0|1.28)^2) = 0.063
\founhhh~(3,3) & $0.37 \pm 0.06 $ & 6.8 & 1.3$^b$ & --- & $2.6 \times 10^{15} $ \\
                         & $0.61 \pm 0.04 $ & 4.7 & 9.5          & ---
                         & --- \\
% Line      Area                   Position            Width             Tpeak
%  1    2.50695E-02 (  0.006)    7.077 (  0.185)    1.300 (  0.000)  1.81163E-02
\founhhh~(2,1) & $0.023 \pm 0.007$ & --- & --- & --- & --- \\
% Line     T ant * Tau           V lsr          delta V            Tau main
%   1  6.810E-2(8.605E-3)   7.22   (7.344E-2)  0.632   (7.966E-3)  0.171   (0.922   )
%   Area=6.81e-2*0.632*SQRT(PI()/4/LN(2))=4.58e-2; 
%   say '0.04581*sqrt((8.605E-3|6.810E-2)^2+(7.966E-3|0.632)^2)' 5.817e-3
% nffts 7.2-1.15 7.2-1.15
%\fivnhhh~(1,1) &  $0.046 \pm 0.007$ & 7.2 & 0.6 & --- & $2.7 \times
%10^{12} $\\
%   1  4.930E-2(1.268E-2)   7.20   (0.110   )  0.853   (0.235   )  0.100   ( 2.14   )
\fivnhhh~(1,1)$^c$ &  $0.045 \pm 0.007$ & 7.2 & 0.85 & --- & $2.6 \times
10^{12} $\\
%nffts 7 8.6
\fivnhhh~(2,2) &  $0.012 \pm 0.008$ & --- & --- & --- & --- \\
\hline
\end{tabular}
\end{center}
Notes: Upper limit for the \founhhh~(2,1) and \fivnhhh~(2,2)
transitions included for  completeness. Molecular column densities are
computed assuming LTE  with temperatures of 10.9 and 14.7~K for
Barnard~1 and NGC~1333,  respectively, as described in the text.
$^a$Uncertainties are from the HFS fit for the \founhhh\
(1,1)--(3,3) lines and computed from the rms in the spectra for the
remaining weak lines and upper limits. $^b$Fixed parameter.  
$^c$The hyperfine structure of \fivnhhh\
has been recently revised by \cite{bethlem08} who give the level
energies with unprecedented accuracy. We have calculated the
corresponding transition strengths using the theory of
\cite{kukolich67} and assuming half-integer values of the quantum
numbers of the levels involved.
\end{table*}

Figure~\ref{fig:spec1} shows spectra of the \founhhh\ and \fivnhhh\
inversion lines toward Barnard~1. The satellite hyperfine lines are
detected with high signal-to-noise ratios for the (1,1) and (2,2)
transitions of \founhhh. Results of hyperfine structure (HFS) fits,
obtained using the IRAM CLASS software package, are shown in
Table~\ref{tab:hfsfits}. The (1,1) and (2,2) \founhhh\ lines have the
same line width of 0.80 \kms, while the (3,3) spectrum is much
broader, indicating the presence of a high-velocity component
originating in warmer gas. Consequently, we fitted the (3,3) line with
two velocity components, fixing the line width of the first component
at the value obtained for the (1,1) and (2,2) lines. All parameters
for the second component were allowed to vary in the fit. The
\fivnhhh~(1,1) line is slightly narrower than the \founhhh~(1,1) and
(2,2) lines, as expected for an optically thin line.
Figure~\ref{fig:spec2} shows the NGC~1333 spectra. The lines are
weaker and broader than in Barnard~1. The (3,3) transition shows
strong blueshifted emission from the molecular outflow driven by
IRAS~4A. The line width derived from the HFS fit to the \fivnhhh\ data
is significantly smaller than the values obtained for \founhhh.
However, the signal to noise ratio in the \fivnhhh\ spectrum is
relatively low, formally 6.2$\sigma$.

We use the rotation diagram technique \citep[see, e.g., eq. 2
of][]{blake87} to derive the excitation temperatures and molecular
column densities. We use the spectroscopic constants and partition
functions for \founhhh\ and \fivnhhh\ from the JPL Molecular
Spectroscopy database (http://spec.jpl.nasa.gov/). The resulting
rotation diagrams are shown in Figures~\ref{fig:spec1} and
~\ref{fig:spec2}, bottom-right panels. The least-squares fit to the
line intensities of the (1,1)--(3,3) data (narrow component) in
Barnard~1 gives a rotation temperature of 12.4~K in (dotted line in
Fig.~\ref{fig:spec1}). However, since the (3,3) spectrum is
contaminated by the high-velocity emission, only the (1,1) transition
is detected for \fivnhhh, and given our interest in the cold gas
component on the line of sight, a more consistent approach is to use
only the (1,1) and (2,2) \founhhh\ data for the temperature
determination. This approach also mitigates possible effects of the
non-LTE ortho-to-para ratio \citep[e.g.,][]{umemoto99} and results in
a slightly lower rotation temperature of $10.9 \pm 0.2$~K (assuming
2.7\% uncertainties for the integrated line intensities, as discussed
in \S~\ref{sec:discussion}; dashed line in Fig.~\ref{fig:spec1},
bottom-right panel). This value is subsequently used to derive the
\founhhh\ and \fivnhhh\ column densities from the (1,1) integrated
line intensities (Table~\ref{tab:hfsfits}). Given the low rotation
temperature, we include the correction for the 2.7~K cosmic background
radiation in the column density computations. This correction has no
effect on the temperature determination, since all the lines observed
have very close rest frequencies. The \nhhh\ temperature and column
density in Barnard~1 we derive here are in good agreement with the
results of \cite{bachiller90} based on observations with the
Effelsberg telescope (12~K and $2.5 \times 10^{15}$~cm$^{-2}$).

For NGC~1333, we derive a rotational temperature of 17.6~K using the
(1,1)--(3,3) lines and $14.7 \pm 0.5$~K using only the (1,1) and (2,2)
transitions (dotted and dashed lines, respectively in
Fig.~\ref{fig:spec2}, bottom-right panel). We use the latter value for
the column density determination.

\section{Discussion}\label{sec:discussion}

The uncertainty of the derived column densities is difficult to
estimate quantitatively. As discussed in \S~\ref{sec:observations},
all lines are observed simultaneously and the relative calibration is
largely determined by the signal-to-noise ratio in the spectra---24
and 6.2$\sigma$ for \fivnhhh~(1,1) in Barnard~1 and NGC~1333,
respectively. The \founhhh/\fivnhhh\ column density ratio, when
computed from the observed intensities of the (1,1) transitions of the
two isotopologues, is insensitive to the exact value of the rotation
temperature used in the calculations, as long as the temperature is
the same for the two isotopologues. The remaining source of
uncertainty is the determination of the \founhhh\ optical depth. The
formal uncertainties for the opacity corrected line intensities given
by the HFS fits are negligible (see Table~\ref{tab:hfsfits}). In order
to independently estimate the uncertainty of the \founhhh (1,1)
integrated line intensities obtained from the HFS fit, we divide the
full Barnard~1 data set into three subsets, 58~min integration time
each, and fit each subset independently. From the three
measurements we derive a 2.7\% uncertainty for the mean which we use
as an estimate of the uncertainty of our final \founhhh\~(1,1)
measurement. Adding in quadrature the corresponding statistical
uncertainty for \fivnhhh\ of 4.2\%, we derive a 3$\sigma$ uncertainty
for the \founhhh/\fivnhhh\ ratio of 15.0\% in Barnard~1. Assuming a
5.2\% uncertainty for the \founhhh\ (1,1) measurement in the shorter
NGC~1333 integration (the Barnard~1 value scaled by the square root of
the integration time ratio), and adding in quadrature a 16\%
uncertainty for the \fivnhhh\ measurement, we derive a 50\%
uncertainty (3$\sigma$) of the \founhhh/\fivnhhh\ ratio in this
source. Our final estimates of the \founhhh/\fivnhhh\ abundance ratio
are thus
%$360 \pm 54$ and $336 \pm 168$ for Barnard 1 and NGC~1333, 
$334 \pm 50$ and $344 \pm 173$ (3$\sigma$) for Barnard 1 and NGC~1333,
respectively.

The $^{14}$N/$^{15}$N fractionation ratios in Barnard~1 and NGC~1333
derived from deuterated ammonia measurements by \cite{gerin08} are
$470^{+170}_{-100}$ and $360^{+260}_{-110}$, respectively. The \nhhh\
and \nhhd\ results are thus consistent, given the \nhhd\ measurement
uncertainties. Owing to the relatively short integration and low SNR
in the \fivnhhh\ spectrum, our result for NGC~1333 is not very
constraining. However, our Barnard~1 data are of excellent quality.
The derived $^{14}$N/$^{15}$N ratio falls in-between the local
ISM/proto-solar value (\about 450, as discussed in \S~1) and the
terrestrial atmospheric ratio of 272. We do not see evidence for the
very high $^{15}$N enhancements, as suggested by the cometary CN
measurements (\about 150).

A comparison of our fractionation measurements with the nitrogen
``superfractionation'' models of \cite{rodgers08b} is of interest. In
their Figure~3, \cite{rodgers08b} present results of time-dependent
models of the evolution of the $^{14}$N/$^{15}$N enhancement ratios in
key species, including ammonia. For a temperature of 10~K, the
gas-phase $^{15}$N fractionation in ammonia can reach a maximum
enhancement of a factor \about 7--10, much higher than the value
we derive for Barnard~1. However, the enhancement factor decreases
with time to a factor of a few \emph{below} the initial isotopic ratio
assumed in the model. The time at which the maximum fractionation is
reached depends strongly on the initial fraction of nitrogen in
molecular form, but is typically of order 1~Myr. The enhancements of
the magnitude that we measure can be present very early on, or a few
tenths of a Myr after the peak enhancement time.

The \cite{rodgers08b} model results depend strongly on the gas
temperature---at 7~K, significantly higher enhancements are found in
the gas-phase ammonia fractionation, as compared to 10~K. The sources
we have studied are somewhat warmer than the model clouds considered
by \cite{rodgers08b}. It would thus be extremely interesting to extend
our measurements to colder clouds, such as LDN~1544 or LDN~134N, with
central temperatures of order 7~K. Our present observations are
exploratory in nature, but show that accurate measurements of nitrogen
fractionation in ammonia in cold ISM sources are clearly feasible
using the GBT. A focused observing program of a larger, carefully
selected sample of prestellar cores with varying temperatures and gas
densities would allow significant progress in our understanding of
nitrogen fractionation in local ISM and its possible implications for
the existing measurements in various solar system reservoirs.

The nitrogen fractionation enhancement levels derived here are well
reproduced by the gas-phase models of Roueff et al. (see, e.g., Fig.~2
of Gerin et al. 2009). For an \hh\ density of $10^5$~\pccm\ and a 10~K
temperature, the gas-phase models predict \founhhh/\fivnhhh~=~389 and
$^{14}$NH$_2$D/$^{15}$NH$_2$D~=~372. The models of Roueff et al. are
steady-state models that focus on the gas-phase processes and
approximate the adsorption of the species on grain surfaces by
reducing the elemental abundances in the gas, as compared to the
dynamical models of Rodgers \& Charnley that explicitely take into
account the mantle formation and gas-grain interactions.

\acknowledgments 
We thank T.~Minter and the GBT operator, D.~Stricklin,
for their assistance during the observations and J.~Braatz for his
advice in setting up the observing scripts. The National Radio
Astronomy Observatory is a facility of the National Science Foundation
operated under cooperative agreement by Associated Universities, Inc.
D.~C.~Lis is supported by the National Science Foundation grant
AST-0540882 to the Caltech Submillimeter Observatory. M.~Gerin and
E.~Roueff acknowledge support from the CNRS-INSU French National
Program PCMI (Physique et Chimie du Milieu Interstellaire).

\end{document}